\documentclass{ifacconf}
\usepackage{graphicx}      % include this line if your document contains figures
\usepackage[square,numbers]{natbib} % For numerical citations
\usepackage{tabularx}
\usepackage{tikz}
\usetikzlibrary{positioning, arrows, patterns} % Add any other libraries you need
\usepackage{pgfplots}
\pgfplotsset{compat=1.18} % Set the version for compatibility, adjust if needed
\usepackage{float}

\usepackage{amsmath}
\usepackage{amsfonts,bm} % Math packages
\usepackage{framed} % Framing content 
\usepackage{nomencl}
\usepackage{float}
\usepackage{optidef}
\usepackage{mathtools}
\usepackage[hyphens]{url}
\DeclareMathSymbol{\sm}{\mathbin}{AMSa}{"39} %short minus
\usepackage{tabularx}
\usepackage{array}

\newcommand{\nox}{$\mathrm{NO_{x}}$}
\newcommand{\cotw}{$\mathrm{CO_{2}}$}

\nomenclature{ML}{Machine Learning}
\nomenclature{MPC}{Model Predictive Control}
\nomenclature{LSTM}{Long-short term memory}
\nomenclature{ICE}{Internal Combustion Engine}
\nomenclature{PPO}{Proximal Policy Optimisation}
\nomenclature{TD3}{Twin Delayed Deep Deterministic Policy Gradients}
\nomenclature{NMPC}{Nonlinear Model Predictive Control}
\nomenclature{LMPC}{Linear Model Predictive Control}
\nomenclature{ARX}{Autoregressive with Extra Input}
\nomenclature{SOI}{Start Of Injection}
\nomenclature{FQ}{Fuel Quantity}
\nomenclature{DPM}{Detailed Physical Model}
\nomenclature{SVM}{Support Vector Machine}
\nomenclature{NOx}{Nitrogen Oxide}
\nomenclature{RNN}{Recurrent Neural Network}
\nomenclature{VGT}{Variable-Geometry Turbocharger }
\nomenclature{OCP}{Optimal Control Problem}
\nomenclature{SQP}{Sequential Quadratic Programming}
\nomenclature{ESM}{Engine Simulation Model}

\begin{document}
\begin{frontmatter}

\title{Safe Reinforcement Learning-based Control for  Hydrogen Diesel Dual-Fuel Engines}
%{Safe Model-based offline Learning based Control for  Hydrogen Diesel Dual-Fuel Engines}

\thanks{Corresponding Author: Vasu Sharma -(e-mail: sharma\_v@mmp.rwth-aachen.de).}

% \author[First]{Armin Norouzi}
% \author[First]{Saeid Shahpouri}
% \author[First]{Mahdi Shahbakhti}
% \author[First]{Charles Robert Koch}
% \address[First]{Mechanical Engineering Department,~University~of~Alberta, Edmonton, Canada}

\author[2]{Vasu Sharma} %113371
\author[2]{Alexander Winkler} %137059
\author[1]{Armin Norouzi} %135760
\author[2]{Jakob Andert}
\author[1]{David Gordon} %129281
\author[3]{Hongsheng Guo}
\address[1]{Department of Mechanical Engineering, University of Alberta, Edmonton, Canada}
\address[2]{Teaching and Research Area Mechatronics in Mobile Propulsion, RWTH Aachen University, Aachen, Germany}
\address[3]{National Research Council, Canada}

This work has been submitted to IFAC for possible publication
\begin{abstract}
The urgent energy transition requirements towards a sustainable future stretch across various industries and are a significant challenge facing humanity. Hydrogen promises a clean, carbon-free future, with the opportunity to integrate with existing solutions in the transportation sector. However, adding hydrogen to existing technologies such as diesel engines requires additional modeling effort. Reinforcement Learning (RL) enables interactive data-driven learning that eliminates the need for mathematical modeling. The algorithms, however, may not be real-time capable and need large amounts of data to work in practice. This paper presents a novel approach which uses offline model learning with RL to demonstrate safe control of a 4.5$\,$L Hydrogen Diesel Dual-Fuel (H2DF) engine. The controllers are demonstrated to be constraint compliant and can leverage a novel state-augmentation approach for sample-efficient learning. The offline policy is subsequently experimentally validated on the real engine where the control algorithm is executed on a Raspberry Pi controller and requires 6 times less computation time compared to online Model Predictive Control (MPC) optimization.
\end{abstract}

\end{frontmatter}
%===============================================================================

%%%%%%%%%%%%%%%%%%%%%%%%%%%%%%%%%%%%%%%%%%%
\section{Introduction}
In the coming decades, global warming could result in a mean global temperature rise of 1.5$^\circ\text{C}$ from preindustrial levels, which would not only have a seismic impact on our way of life, but also jeopardize sensitive ecosystems worldwide~\citep{IntergovernmentalPanelonClimateChange.2022}. Despite major advances, the transportation sector continues to be a significant contributor towards green house gas emissions, with estimated contributions from 16 to 23 percent ~\citep{Crippa.2023}. Within the transportation sector, the heavy-freight industry disproportionately contributes to roughly 35 percent of the total carbon dioxide~(\cotw) emissions, despite having a total vehicle footprint of less than 8 percent through buses and trucks \citep{TrackingCleanEnergyProgress2023IEA.2023}.

Owing to their high thermal efficiency and low maintenance, diesel engines have been a cornerstone of the heavy-duty transportation sector. However, due to ever stringent emission regulations worldwide, a cleaner alternative has been sought by the sector. Hydrogen promises a carbon-free alternative which can be used
in conjunction with electrification to offer a smoother energy transition for the
transportation sector. When combined with conventional diesel engines, hydrogen replaces diesel from the engine in the combustion energy share, thus promoting lower particulate and~\cotw~emissions. However, complex combustion characteristics can limit the hydrogen fraction and cause abnormal combustion at higher loads, and therefore require complex modeling and calibration. Additionally, higher operating temperatures can lead to increased nitrogen oxide (NOx) formation \citep{Tsujimura.2017}.

With the prospect to synthesize controllers directly from data, Reinforcement Learning (RL) offers an alternative approach. In the last few years, RL has been used to control the variable geometry turbocharger (VGT) vane position along with the injection timing of a diesel engine, to maximize the engine break torque \citep{Malikopoulos.2009}, transient boost pressure tracking by controlling the vane position of a VGT-equipped diesel engine \citep{Hu.2019} and fuel consumption minimization among other use cases \citep{Egan.2023}. Virtualized development platforms have also been shown to be effective in saving online training efforts \citep{Koch.2023b, Koch.2023, Badalian.2024}.

However, results from the current literature utilize high-fidelity simulation training models, which may not always be available or, when available, are not real-time capable. This also leads to challenges when bringing the learned policies in the simulated environment to the real world \citep{Li.2024}. 
Moreover, despite the success in simulated environments, only a few multi-objective real-world use cases can be found in the literature \citep{Norouzi.2023}. The control actions are also often low-dimensional and build on existing engine applications instead of controlling the entire combustion process itself.

This paper aims to tackle the aforementioned challenges by combining state-of-the-art deep learning models to enable sample-efficient reinforcement learning for a hydrogen diesel engine. The methodology is demonstrated to work for both on-policy and off-policy agents which are trained in a simulated environment. The trained policies are then shown to control the actual process. To the best of the authors' knowledge, this paper presents the first study to use this technique for real-time control of a hydrogen diesel engine.

Based on the current literature, this paper has the following contributions:

\begin{enumerate}
    \item Deep learning-based offline model learning for RL design
        \begin{enumerate}
    \item A Gated Recurrent Unit~(GRU)-based encoder-decoder model is developed to capture and represent the engine dynamics and emissions 

        \end{enumerate}
    \item Offline reinforcement learning based control 
        \begin{enumerate}
    \item RL agents are trained on the learned model to minimize engine-out emissions and fuel consumption while maintaining the same output load tracking performance.
    \item Developed model-based offline-RL controller is demonstrated to reduce the computational time of optimization
    \end{enumerate}
\end{enumerate}

%%%%%%%%%%%%%%%%%%%%%%%%%%%%%%%%%%%%%%%%%%%
\section{Methodology}

\begin{figure}
    \centering
    \includegraphics[width = 0.46\textwidth]{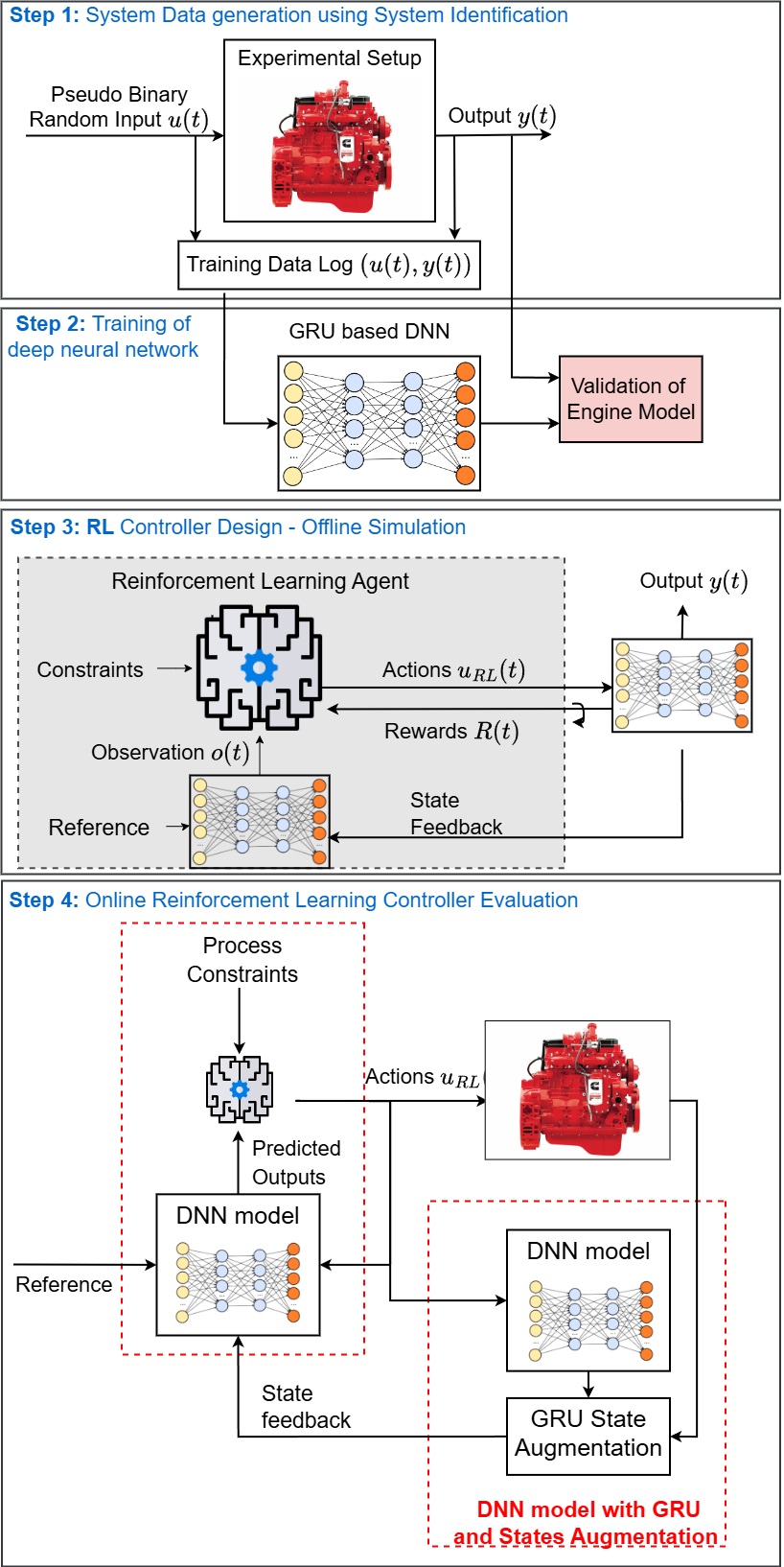}
    \caption{Modeling and controller design procedure based on deep reinforcement learning }
    \label{fig:offlinemodeling}
\end{figure}

In this paper, a four-step methodology combining deep neural networks (DNN) and RL to control a hydrogen diesel engine is presented. An overview of the methodology is depicted in Fig.~\ref{fig:offlinemodeling}. Step 1 focuses on system identification using random actuations applied to the engine, and recording the outputs. Then, a GRU-based encoder-decoder style model is learned to represent the plant dynamics through deep learning. This is subsequently used in conjunction with a reward function and state augmentations to learn load tracking and constraint-compliant control through offline model-based reinforcement learning. The effects of state feedback are studied on training and performance for Proximal Policy Optimization (PPO) and Twin Delayed Deep Deterministic Policy Gradients (TD3) agents. Finally, the learned neural network policies are executed on the real engine using a Raspberry Pi controller.

%%%%%%%%%%%%%%%%%%%%%%%%%%%%%%%%%%%%%%%%%%%
\section{GRU Encoder-Decoder based DNN Engine Model}
\subsection{Experimental Setup Details}
One cylinder of a Cummins QSB 4.5$\,$L 4-cylinder (EPA Offroad Tier 3 certified) engine is modified for port-injected hydrogen to allow for H2DF operation. Hydrogen is injected into the intake runner at the beginning of the intake stroke. Valve timing, combustion chamber geometry, and diesel injectors are stock. Diesel rail pressure is kept at a constant 970$\,$bar for all tests. The production engine control unit (ECU) is replaced with a dSPACE MABX~2 (MABX) and RapidPro power electronics. The MABX contains a Field Programmable Gate Array (FPGA) which is used for real-time calculation of the combustion parameters used in this work, including Maximum Pressure Rise Rate (MPRR) and Indicated Mean Effective Pressure~(IMEP). The MABX allows for fully flexible control of the diesel injection, hydrogen injection, and diesel pressure. The MABX is then connected to a Raspberry Pi 400 where the developed RL control strategies are run. More details outlining the experimental setup can be found in~\citep{Gordon.2022}.
 
\subsection{Offline Model Learning}

\begin{figure}
    \centering
    \includegraphics[trim = 10 370 40 5, clip, width = 0.45\textwidth]{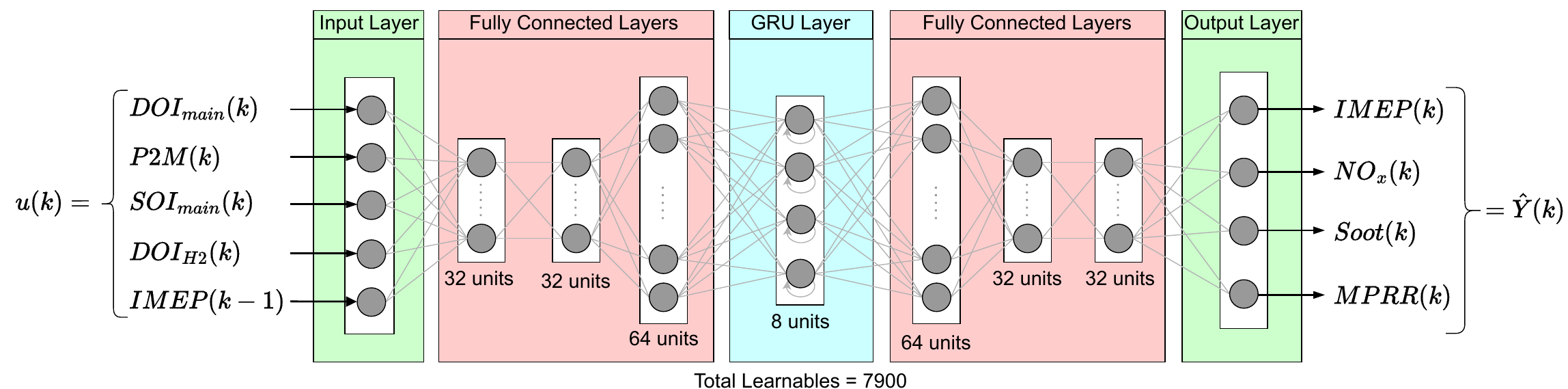}
    \caption{Deep Neural Network Engine Plant Model. GRU: gated recurrent unit, DOI: duration of injection, P2M: pre to main injections, SOI: start of injection, IMEP: indicated mean effective pressure, MPRR: maximum pressure rise rate.}
    \label{fig:DNNmodel}
\end{figure}

To model the engine dynamics, performance, and emissions, a DNN model with six fully connected layers and one recurrent unit layer was developed as shown in Fig.~\ref{fig:DNNmodel}.
The model structure has proven to be effective in similar works in \citep{Gordon.2022, Norouzi.2023, Gordon.2024}. 
The recurrent units within the DNN are GRUs, which can be described by
\begin{subequations}\label{eq:GRU_equ}
   \begin{align}   
    z(t) &= \sigma\left(W_{h,z}^\top h(t-1) + W_{u,z}^\top u(t) + b_{z}\right), \\
    r(t) &= \sigma\left(W_{h,r}^\top h(t-1) + W_{u,r}^\top u(t) + b_{r}\right) ,\\
    \tilde{h}(t) &= \tanh\left(W_{u,\tilde{h}}^T u(t) + W_{h,\tilde{h}}^\top (r(t) \odot h(t-1)) + b_{\tilde{h}}\right),\\
    h(t) &= (1-z(t)) \odot h(t-1) + z(t) \odot \tilde{h}(t).
   \end{align}
\end{subequations}
where $u(t)$ is the input vector to the unit, $z(t)$ and $r(t)$ are the update gate vector and
reset gate vector, respectively. $h(t)$ is the output vector of the unit, $\tilde{h}(t)$ is the candidate activation vector, while $W$ and $b$
are the parameter weights and biases to be learned during training. GRU offers the advantage of achieving similar performance in time-series prediction as Long Short-Term Memory units, while using fewer learnable parameters and requiring fewer operations, making them more efficient during training and execution.
\subsubsection{Plant Training: }As shown in Fig.~\ref{fig:DNNmodel}, the inputs $u(t)$ and outputs $y(t)$ of the engine plant model can be denoted as 
\begin{align}
u(t) &=
\begin{bmatrix}
u_{\text{DOI,fuel}}(t) & u_{\text{P2M}}(t) & u_{\text{SOI,fuel}}(t) \\ 
u_{\text{DOI,H2}}(t) & y_{\text{IMEP}}(t - 1)
\end{bmatrix}^T, \notag \\
y(t) &=
\begin{bmatrix}
y_{\text{IMEP}}(t) & y_{\text{NOx}}(t) & y_{\text{Soot}}(t) & y_{\text{MPRR}}(t)
\end{bmatrix}^T.
\end{align}
The duration of injection (DOI) of the two fuels, diesel (Main) and hydrogen (H2), the time between the pre and the main diesel injection (P2M), and the starting crank angle degree (CAD) of the main diesel injection (SOI) are the control variables of the engine and the DNN model inputs. The engine and DNN model outputs are the IMEP which represents the engine load, nitrogen oxide emissions \nox, particulate emissions(Soot) and MPRR. 

A dataset for the DNN training of 85,000 consecutive cycles was collected from the engine utilizing a pseudo-random binary input generator to train the DNN. The training parameters are shown in Tab.~\ref{tab:TrainingSettings}, with an 80/15/5 split between training, validation, and unseen test data.
\subsubsection{Results: }
\begin{figure}[htbp]
\centering
\includegraphics[ width = 0.48\textwidth]{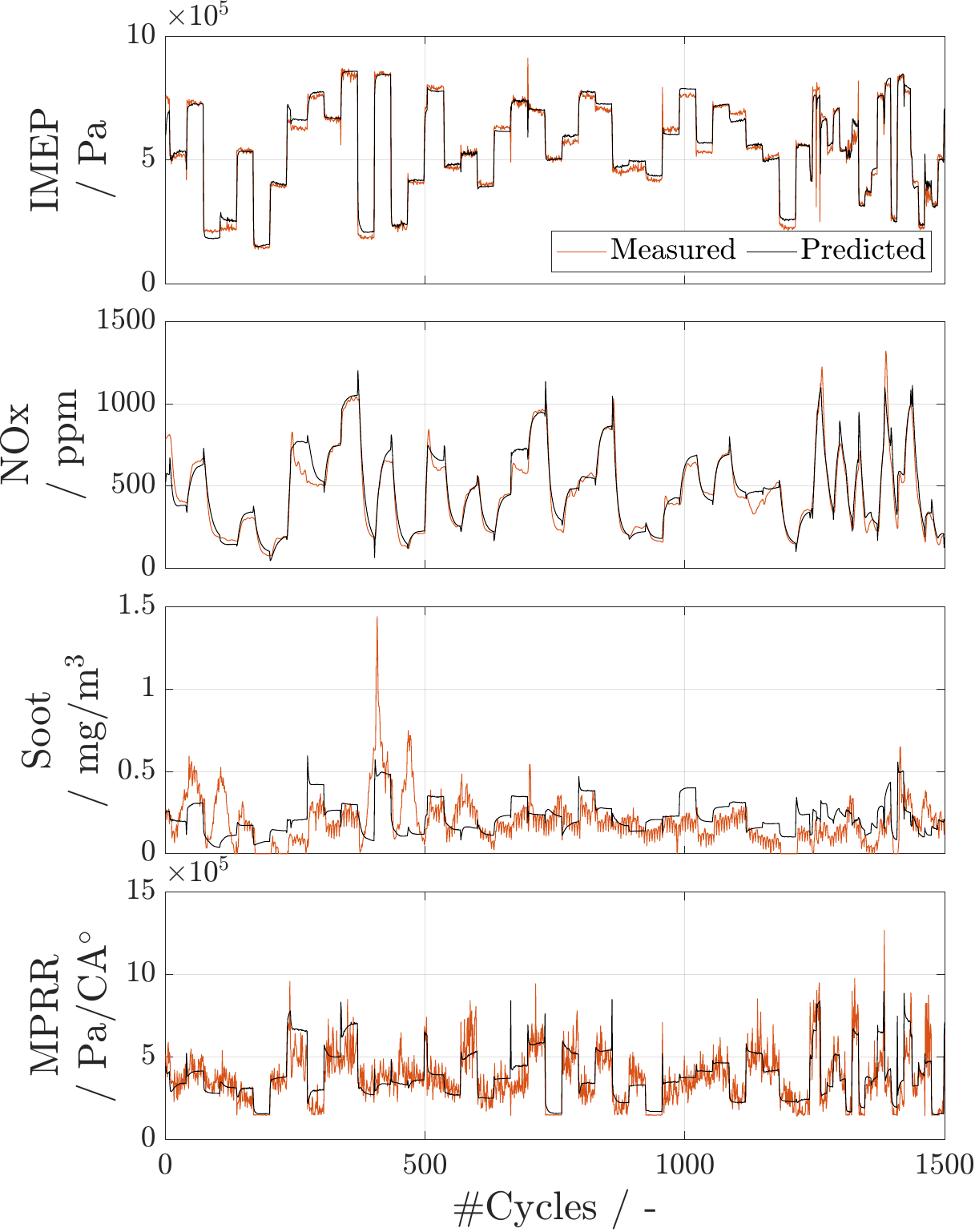}
\caption{GRU-based network outputs on the test set}
\label{fig:NNTestResults}
\end{figure}

The resulting accuracy of the DNN model can be considered sufficient, with the Root Mean Squared Percentage Error (RMSPE) being under 4.6\% for all four prediction outputs on the unseen test dataset. Fig.~\ref{fig:NNTestResults} shows the prediction on the unseen test dataset time-series.
\begin{table}[htbp]
    \centering
    \renewcommand{\arraystretch}{1.3}
    \caption{Offline plant learning hyperparameters}
    \begin{tabular*}{0.45\textwidth}{l @{\extracolsep{\fill}} @{\hspace{1em}} | c}
        \hline
        \hline
        \textbf{Parameter} & \textbf{Value} \\
        \hline
        \hline
        Maximum Epochs              & 1000 \\
        Batch Size                  & 512 \\
        Gradient Threshold          & 1 \\
        L2 Regularization           & 0.01 \\
        Learning Rate Decay         & Piecewise at 250 Epochs \\
        Initial Learning Rate       & 0.0005 \\
        Learning Rate Drop Factor   & 0.75 \\
        Validation Patience         & 3 \\
        \hline
    \end{tabular*}
    \label{tab:TrainingSettings}
\end{table}

\section{Offline Reinforcement Learning}
\subsection{Reward Modeling with Additive Noise}
The reward formulation term (Eq. \ref{eq:RewardForm}) consists of multiple 
penalty terms, namely: control input minimization term \( r \), output deviation term \( q \), reward staging term \(R^{-}\), constraint violation penalty term \( S \). The reward is calculated by adding all penalty terms and taking the negative absolute value (Eq.~\ref{eq:RewardForm}). The formulation is designed for multi-objective optimization, as commonly seen in model predictive control (MPC)~\citep{Gordon.2022}.

\begin{figure}
    \centering
    \includegraphics[width = 0.46\textwidth]{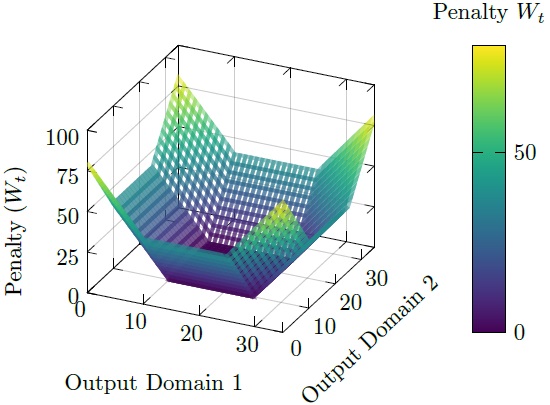}
    \caption{Constraint violation penalty visualized over multiple output domains.}
    \label{fig:ConstraintViolationPenalty3D}
\end{figure}

The output deviation term rewards the agent for strict tracking to the required IMEP load reference, while minimizing the emissions (\nox~and Soot) and ensuring engine durability (MPRR). In contrast, the control minimization term penalizes the agent for larger control actions during tracking. The reward staging term (Eq. \ref{eq:rewardstaging}) exponentially rewards the agent as it matches the tracking target, where the extra gradient information aims to mitigate the diminishing gradients from the least-square term ($q_1$).
\begin{equation}
\scalebox{0.8}{$  % Adjust scaling factor (0.9) as needed
\sum_{t=0}^{N} R_t = - \left( 
\begin{aligned}
& \alpha \cdot\sum_{t=0}^{N} \Bigg( \|
 \underbrace{\left( r_{\text{IMEP}_t} - y_{\text{IMEP}_t} \right)^2}_{q_1 \text{ Load Tracking}} +
 \underbrace{\left( y_{\text{MPRR}_t} \right)^2}_{q_2 \text{ Combustion Noise}} \\
& \quad + 
 \underbrace{\left( y_{\text{NOx}_t} \right)^2 + \left( y_{\text{Soot}_t} \right)^2}_{q_3 ~\text{Emissions Minimization}} \Bigg\|_Q \Bigg) \\
& + \beta \cdot \sum_{t=0}^{N}  \Bigg( \|
 \underbrace{\left( u_{\text{H2DOI}_{\text{main}, t}} \right)^2 + \left( u_{\text{DOI}_{\text{Main}, t}} \right)^2}_{r~\text{Control ...}} \\
& \quad + \underbrace{\left( u_{\text{P2M}_t} \right)^2 + \left( u_{\text{SOI}_{\text{Main}, t}} \right)^2}_{\text{...Effort Minimization}} \Bigg\|_{\tilde{R}} \Bigg) \\
& + \sum_{t=0}^{N} \Bigg( 
\underbrace{\|R^{-}_{t}\|}_{R^{-} \text{ Reward Staging}} + \zeta\cdot\underbrace{\|W_t\|}_{W \text{ Constraint Violation}}
\Bigg)
\end{aligned}
\right)$}
\label{eq:RewardForm}
\end{equation}
\begin{equation}
R^{-}_{t} = -0.0025 \times 10^{\left\lfloor 5 - \log_{10}(\delta \text{IMEP}_t) \right\rfloor},
\label{eq:rewardstaging}
\end{equation}
where $\delta \text{IMEP}_t$ refers to $\left | r_{\text{IMEP}_t} - y_{\text{IMEP}_t}\right|$. Finally, for the agent to discover a safe operating region in the output space, the constraint violation term $S$ signals safety violations caused by the agent's actions, via large penalties $W_t$. Safety bounds are pre-selected based on process knowledge and any violations outside the safe polytope result in an $L_1$-style  penalty. The formulation allows soft-safety to be directly included in the optimization objective of the problem and builds on previous work \citep{Norouzi.2023}. The resultant penalty polytope is visualized in Fig.~\ref{fig:ConstraintViolationPenalty3D} for two arbitrary output domains.

The terms in \( r \) and \( q \) are weighted with weight matrices \( \tilde{R} \) and \( Q \). Moreover, three constants $\alpha, \beta, \zeta$ are needed for relative weighting of the terms between \( r \), \( q \) and \( S \), thus allowing large penalties for constraint violations and output deviations.

\subsection{State Augmentation with Perturbed Rewards}

The trained plant model is a deterministic mapping ($\text{f}_{\text{GRU-NN}}$) from the controls to the output space. While it matches the real process closely in simulations, the experimentally measured signals could still be noisy. Additive Gaussian noise is thus added to the outputs during training to better represent the noisy signals from the actual process. This also perturbs the reward signals, which may help the optimizer from getting stuck in local minima.

The applied RL controls ($u_{RL}$) result in an internal state ($\hat{h}_{\text{GRU}}(t)$) change for the DNN model-based environment. These updated states can also be accessed by propagating the estimated outputs of the network (Eq. \ref{eq:stateaugmentation}). This lookahead during learning informs the RL agents about the current system state. Based on the accuracy of the model and the reward formulation, the agent can choose to use or reject this information with gradient updates. The state feedback also introduces a multi-step temporal feature into the observation model.
    \begin{equation}
\begin{aligned}
    \text{f}_{\text{GRU-NN}}(u_{RL}(t), \tilde{y}_{\text{IMEP}}(t-1)) \to \hat{y}(t), \\
    \hat{y}(t)= \left[ \hat{y}_{\text{IMEP}}(t), \hat{y}_{\text{NOx}}(t), \hat{y}_{\text{Soot}}(t), \hat{y}_{\text{MPRR}}(t) \right]^T, \\
    \tilde{y}(t) \gets \hat{y}(t) + \epsilon, \quad \epsilon \sim \mathcal{N}(0, \tilde{\sigma}), \\
    \text{f}_{\text{GRU-NN}}(u_{RL}(t), \tilde{y}_{\text{IMEP}}(t-1)) \to \hat{h}_{\text{GRU}}(t).
\end{aligned}
\label{eq:stateaugmentation}
\end{equation}

\subsection{RL Training}
\begin{figure}[htbp]
% Adjust spacing before the figure
\centering
\includegraphics[width=0.45\textwidth]{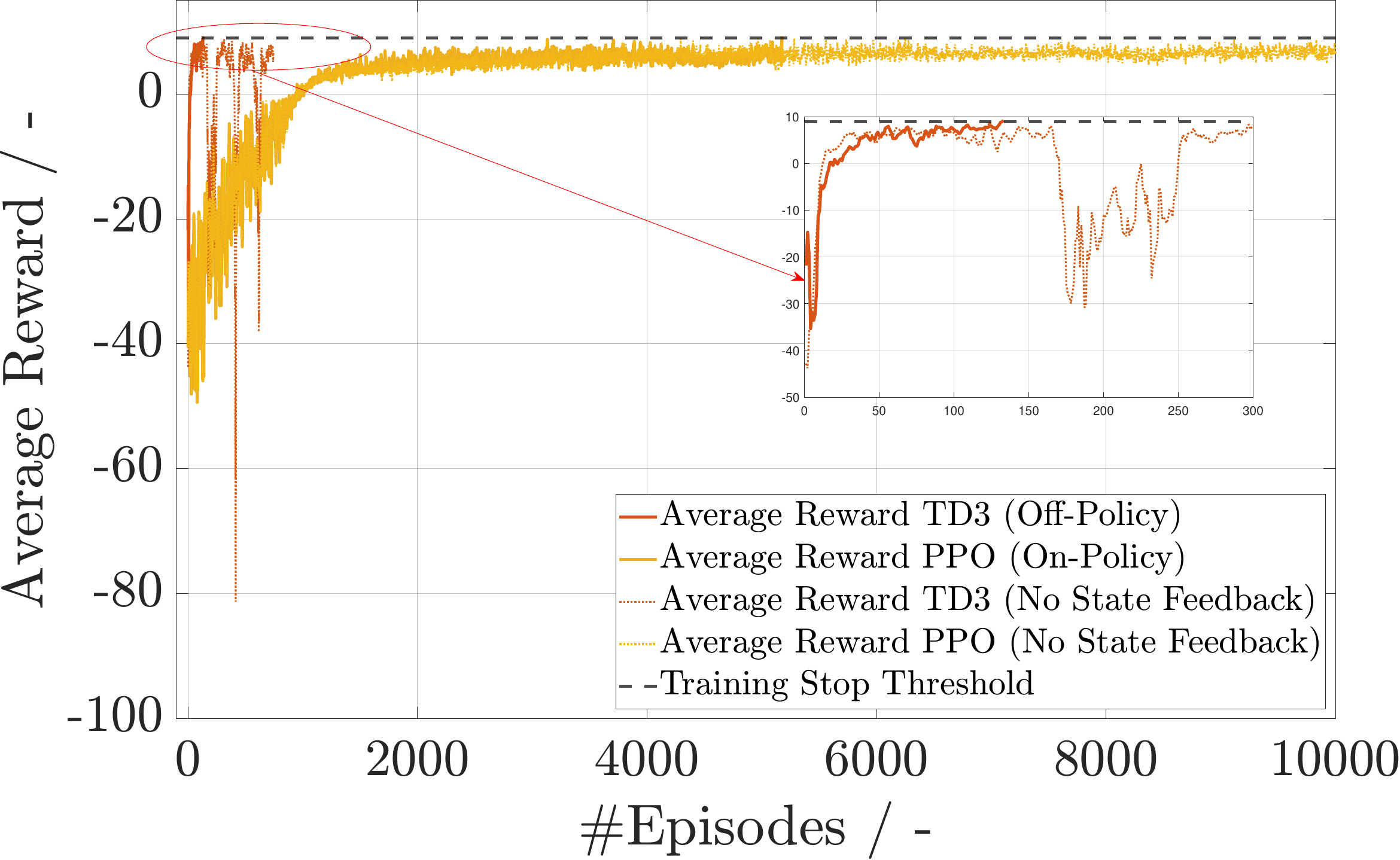}
\caption{RL training for various agents}
\label{fig:Td3vsPPO}
\end{figure}

The RL agent observes the environment states $o(t)$, including the outputs from the last applied controls, the current reference load, and the augmented plant states (Eq.~\ref{eq:ot}, Eq. \ref{eq:otdefinitions}) to decide on the current actions $u_{\text{RL}}$. Thus $o(t)$ can be defined as:
\begin{align}
o(t) &= 
\begin{bmatrix} 
\tilde{y}(t\!-\!1) & {\Delta \tilde y}(t\!-\!1) & r_{\text{load}} & {h}_{\text{GRU}}(t) 
\end{bmatrix}^T
\label{eq:ot}
\end{align}
\begin{align}
\begin{split}
    \tilde{y}(t\!-\!1) &= \begin{bmatrix} y_{\text{IMEP}}(t\!-\!1) & y_{\text{NOx}}(t\!-\!1) & y_{\text{MPRR}}(t\!-\!1) \end{bmatrix}^T, \\
    {\Delta \tilde y}(t\!-\!1) &= \begin{bmatrix} \Delta y_{\text{IMEP}}(t\!-\!1) & \Delta y_{\text{NOx}}(t\!-\!1) \end{bmatrix}^T, \\
    r_{\text{load}} &= \begin{bmatrix} r_{\text{IMEP,ref}}(t) & r_{\text{IMEP,ref}}(t\!-\!1) & r_{\text{IMEP,Err}}(t\!-\!1) \end{bmatrix}^T, \\
    {h}_{\text{GRU}}(t) &= \hat{h}_{\text{GRU}}(t).
\end{split}
\label{eq:otdefinitions}
\end{align}
\begin{equation}
u_\text{RL}(t) = 
\begin{bmatrix}
u_{\text{DOI,fuel}}(t) & u_{\text{P2M}}(t) & u_{\text{SOI,fuel}}(t) & u_{\text{DOI,H2}}(t)\end{bmatrix}^T.
\end{equation}
Using the previously elaborated reward model, off-policy and on-policy algorithms are trained in an episodic setting. Random trajectories are generated per episode for a total episodic length of 625 time-steps. Twin Delayed Deep Deterministic Policy Gradients (TD3) is an off-policy, actor critic-style RL algorithm that uses randomly sampled experiences from an experience buffer to train a deterministic policy \citep{Fujimoto.2018}. For improved stability, delayed target updates with policy smoothing and two critic networks are used. Proximal Policy Optimization (PPO) is an on-policy, advantage actor critic algorithm that uses a clipped surrogate objective \citep{Schulman.2017}. Both of these are used in this work. The relevant training parameters are included in Tab.~\ref{tab:OfflineTrainingHyperparams}.
\begin{table}[htbp]
\centering
\renewcommand{\arraystretch}{1.3}
\caption{RL training hyperparameters}
\begin{tabular*}{0.45\textwidth}{l @{\extracolsep{\fill}} @{\hspace{2em}} | c}
\hline \hline
\textbf{Parameter} & \textbf{Value} \\
\hline \hline
Observation Space Size &            16 \\
Action Space Size &          4 \\
Training Stop Threshold &       9 \\
Optimizer &         Adam \\
Learning Rate &         0.0001 \\
Gradient Threshold &        2 \\
Discount Factor &            0.99 \\
Experience Buffer Size &            100,000 \\
Sample Time &           0.08 \\
\hline
\end{tabular*}
\label{tab:OfflineTrainingHyperparams}
\end{table}
The training plots presented in Fig.~\ref{fig:Td3vsPPO} highlight the sample efficiency of the state augmented off-policy method. Both state augmented cases not only reach higher reward values, but also do so much faster, compared to their non state augmented counterparts.
\subsection{RL Validation}
\begin{figure}[htbp]
\centering
\includegraphics[width=0.49\textwidth]{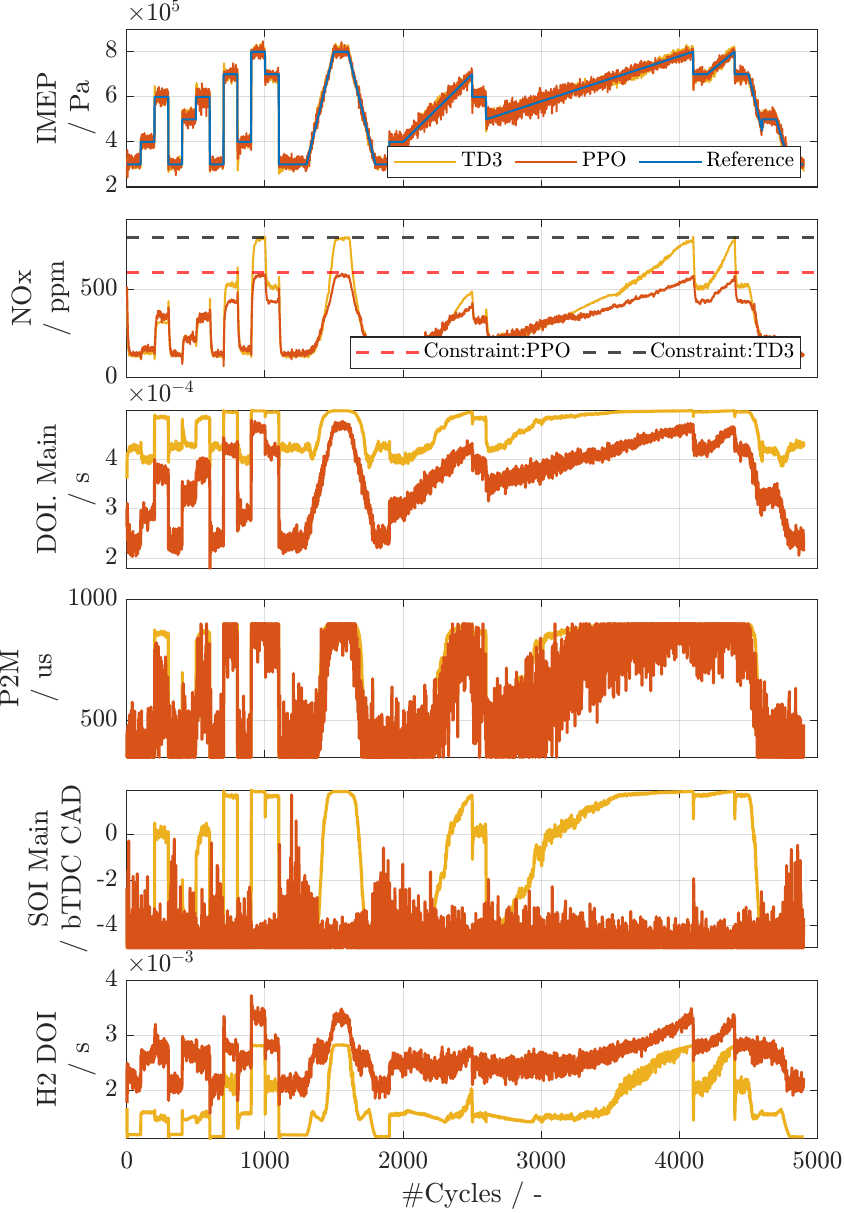}
\caption{Constraint compliant load tracking: PPO and TD3}
\label{fig:ValResults}
\end{figure}

The trained RL agents are simulated over a random virtual trajectory with a network in the loop simulation. The agents observe the reference IMEP load to decide on control actions. To demonstrate the robustness of the proposed method, the validation trajectory is approximately 8 times longer than training trajectories with multiple load changes and ramps. The results are included in Fig.~\ref{fig:ValResults}. 
Both methods demonstrate excellent tracking behavior, highlighting good credit assignment and generalization of the agents. Noteworthy is the difference in control action chosen by the different agents through the various training constraints. The choice of fuel is particularly of interest, where the PPO agent is able to learn to deploy more hydrogen for the same output load, whereas the TD3 agent prefers to use more diesel, since it has a higher NOx tolerance during training. 

Another key point is the difference in behavior of the SOI. While the TD3 agent chooses to almost replicate the requested trajectory, the PPO agent does not really change the SOI trajectory much. This behavior suggests that the engine is not particularly sensitive to SOI changes.

\section{Real-Time Implementation}
\begin{figure}[htbp]
\centering
\includegraphics[ width = 0.49\textwidth]{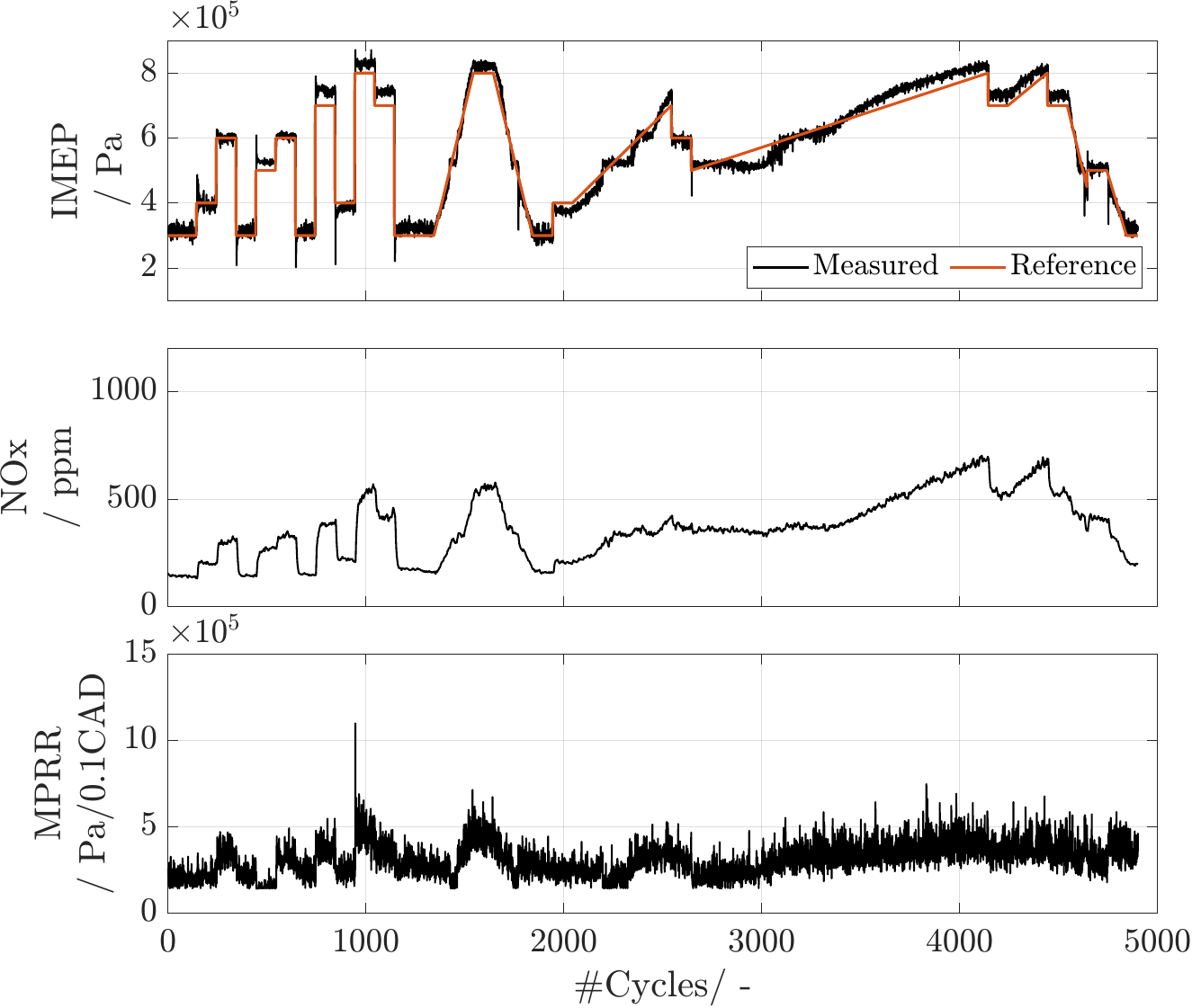}
\caption{Real-time policy validation results on the actual engine using a neural cascaded controller}
\label{fig:RealTimeimplementation}
\end{figure}

The neural network policies from the trained agents are  validated on the actual engine using an ARM Cortex-A72  64-bit Raspberry Pi 400. MATLAB's Raspberry Pi interface, Coder, and the Deep Learning toolbox are used to generate executable C code. The  Pi 400 receives the current engine states from the MABX, along with the desired IMEP via User Datagram Protocol (UDP) and evaluates the policy network to get control actions. Based on the performance of various networks, a cascaded policy selection controller is developed. The total turnaround time is 1$\,$ms which is 6 times faster than a comparative GRU-based MPC. The validation results are included in Fig.~\ref{fig:RealTimeimplementation}.
\section{Summary and Conclusions}
This paper presents the integration of reinforcement learning with offline plant learning for controller implementation for a hydrogen diesel engine application. First, a deep learning-based Gated Recurrent Unit (GRU) encoder-decoder model is trained based on real engine data. The learned plant model is then used to train off-policy (TD3) and on-policy (PPO) agents, with a constraint aware reward model and state augmentation. The trained policies are then executed on a Raspberry Pi to control the real engine, which demonstrates a significant reduction in computation time.

The paper presents a novel end-to-end data-driven approach to control a real process. The modular approach to plant learning and control can leverage the latest developments in deep learning and reinforcement learning in the future. The developed toolchain does not need any special hardware for real-time execution of the learned policies. Real-time capable online fine-tuning could further improve the presented results. The method could also be applied to carbon-free fuels for further emission reduction.

\section*{Acknowledgments}
The author(s) disclosed receipt of the following financial support for the research, authorship, and/or publication of this article: The research was carried out as part of the NRC-MITACS-RWTH Globalink Research Award Program, which is funded by the National Research Council Canada (NRC), MITACS, and Rheinisch-Westfälische Technische Hochschule (RWTH) Aachen University.
\bibliography{Bib_MA}

\begin{thebibliography}{16}
\providecommand{\natexlab}[1]{#1}
\providecommand{\url}[1]{\texttt{#1}}
\providecommand{\urlprefix}{URL }
\expandafter\ifx\csname urlstyle\endcsname\relax
  \providecommand{\doi}[1]{doi:\discretionary{}{}{}#1}\else
  \providecommand{\doi}{doi:\discretionary{}{}{}\begingroup \urlstyle{rm}\Url}\fi

\bibitem[{Badalian et~al.(2024)Badalian, Koch, Brinkmann, Picerno, Wegener, Lee, and Andert}]{Badalian.2024}
Badalian, K., Koch, L., Brinkmann, T., Picerno, M., Wegener, M., Lee, S.Y., and Andert, J. (2024).
\newblock Lexci: A framework for reinforcement learning with embedded systems.
\newblock \emph{Applied Intelligence}, 54(17-18), 8384--8398.
\newblock \doi{10.1007/s10489-024-05573-0}.

\bibitem[{Crippa et~al.(2023)Crippa, Guizzardi, Schaaf, Monforti-Ferrario, Quadrelli, {Risquez Martin}, Rossi, Vignati, Muntean, {Brandao De Melo}, Oom, Pagani, Banja, Taghavi-Moharamli, K{\"o}ykk{\"a}, Grassi, Branco, and San-Miguel}]{Crippa.2023}
Crippa, M., Guizzardi, D., Schaaf, E., Monforti-Ferrario, F., Quadrelli, R., {Risquez Martin}, A., Rossi, S., Vignati, E., Muntean, M., {Brandao De Melo}, J., Oom, D., Pagani, F., Banja, M., Taghavi-Moharamli, P., K{\"o}ykk{\"a}, J., Grassi, G., Branco, A., and San-Miguel, J. (2023).
\newblock \emph{GHG emissions of all world countries: 2023}, volume 31658 of \emph{EUR}.
\newblock {Publications Office of the European Union}, Luxembourg.
\newblock \doi{10.2760/953332}.

\bibitem[{Egan et~al.(2023)Egan, Zhu, and Prucka}]{Egan.2023}
Egan, D., Zhu, Q., and Prucka, R. (2023).
\newblock A review of reinforcement learning-based powertrain controllers: Effects of agent selection for mixed-continuity control and reward formulation.
\newblock \emph{Energies}, 16(8), 3450.
\newblock \doi{10.3390/en16083450}.

\bibitem[{Fujimoto et~al.(2018)Fujimoto, {van Hoof}, and Meger}]{Fujimoto.2018}
Fujimoto, S., {van Hoof}, H., and Meger, D. (2018).
\newblock Addressing function approximation error in actor-critic methods.
\newblock \urlprefix\url{http://arxiv.org/pdf/1802.09477}.

\bibitem[{Gordon et~al.(2022)Gordon, Norouzi, Winkler, McNally, Nuss, Abel, Shahbakhti, Andert, and Koch}]{Gordon.2022}
Gordon, D.C., Norouzi, A., Winkler, A., McNally, J., Nuss, E., Abel, D., Shahbakhti, M., Andert, J., and Koch, C.R. (2022).
\newblock End-to-end deep neural network based nonlinear model predictive control: Experimental implementation on diesel engine emission control.
\newblock \emph{Energies}, 15(24), 9335.
\newblock \doi{10.3390/en15249335}.

\bibitem[{Gordon et~al.(2024)Gordon, Winkler, Bedei, Schaber, Pischinger, Andert, and Koch}]{Gordon.2024}
Gordon, D.C., Winkler, A., Bedei, J., Schaber, P., Pischinger, S., Andert, J., and Koch, C.R. (2024).
\newblock Introducing a deep neural network-based model predictive control framework for rapid controller implementation.
\newblock In \emph{2024 American Control Conference (ACC)}, 5232--5237.
\newblock \doi{10.23919/ACC60939.2024.10644830}.

\bibitem[{Hu et~al.(2019)Hu, Yang, Li, Li, and Bai}]{Hu.2019}
Hu, B., Yang, J., Li, J., Li, S., and Bai, H. (2019).
\newblock Intelligent control strategy for transient response of a variable geometry turbocharger system based on deep reinforcement learning.
\newblock \emph{Processes}, 7(9), 601.
\newblock \doi{10.3390/pr7090601}.

\bibitem[{{Intergovernmental Panel on Climate Change}(2022)}]{IntergovernmentalPanelonClimateChange.2022}
{Intergovernmental Panel on Climate Change} (2022).
\newblock Summary for policymakers.
\newblock In \emph{Global Warming of 1.5°C: IPCC Special Report on Impacts of Global Warming of 1.5°C above Pre-industrial Levels in Context of Strengthening Response to Climate Change, Sustainable Development, and Efforts to Eradicate Poverty}, 1--24. {Cambridge University Press}.
\newblock \doi{10.1017/9781009157940.001}.

\bibitem[{Koch et~al.(2023{\natexlab{a}})Koch, Brinkmann, Wegener, Badalian, and Andert}]{Koch.2023b}
Koch, L., Brinkmann, T., Wegener, M., Badalian, K., and Andert, J. (2023{\natexlab{a}}).
\newblock Adaptive traffic light control with deep reinforcement learning: An evaluation of traffic flow and energy consumption.
\newblock \emph{IEEE Transactions on Intelligent Transportation Systems}, 24(12), 15066--15076.
\newblock \doi{10.1109/TITS.2023.3305548}.

\bibitem[{Koch et~al.(2023{\natexlab{b}})Koch, Picerno, Badalian, Lee, and Andert}]{Koch.2023}
Koch, L., Picerno, M., Badalian, K., Lee, S.Y., and Andert, J. (2023{\natexlab{b}}).
\newblock Automated function development for emission control with deep reinforcement learning.
\newblock \emph{Engineering Applications of Artificial Intelligence}, 117, 105477.
\newblock \doi{10.1016/j.engappai.2022.105477}.

\bibitem[{Li et~al.(2024)Li, He, Khajepour, Chen, Huo, and Wang}]{Li.2024}
Li, Y., He, H., Khajepour, A., Chen, Y., Huo, W., and Wang, H. (2024).
\newblock Deep reinforcement learning for intelligent energy management systems of hybrid-electric powertrains: Recent advances, open issues, and prospects.
\newblock \emph{IEEE Transactions on Transportation Electrification}, 1.
\newblock \doi{10.1109/TTE.2024.3377809}.

\bibitem[{Malikopoulos et~al.(2009)Malikopoulos, Papalambros, and Assanis}]{Malikopoulos.2009}
Malikopoulos, A.A., Papalambros, P.Y., and Assanis, D.N. (2009).
\newblock A learning algorithm for optimal internal combustion engine calibration in real time.
\newblock \emph{ASME 2007 International Design Engineering Technical Conferences and Computers and Information in Engineering Conference}, 91--100.
\newblock \doi{10.1115/DETC2007-34718}.

\bibitem[{Norouzi et~al.(2023)Norouzi, Shahpouri, Gordon, Shahbakhti, and Koch}]{Norouzi.2023}
Norouzi, A., Shahpouri, S., Gordon, D., Shahbakhti, M., and Koch, C.R. (2023).
\newblock Safe deep reinforcement learning in diesel engine emission control.
\newblock \emph{Proceedings of the Institution of Mechanical Engineers. Part I, Journal of systems and control engineering}, 237(8), 1440--1453.
\newblock \doi{10.1177/09596518231153445}.

\bibitem[{Schulman et~al.(2017)Schulman, Wolski, Dhariwal, Radford, and Klimov}]{Schulman.2017}
Schulman, J., Wolski, F., Dhariwal, P., Radford, A., and Klimov, O. (2017).
\newblock Proximal policy optimization algorithms.
\newblock \urlprefix\url{http://arxiv.org/pdf/1707.06347}.

\bibitem[{{Tracking Clean Energy Progress 2023, IEA}(2023)}]{TrackingCleanEnergyProgress2023IEA.2023}
{Tracking Clean Energy Progress 2023, IEA} (2023).
\newblock Iea (2023), tracking clean energy progress 2023, iea, paris https://www.iea.org/reports/tracking-clean-energy-progress-2023, licence: Cc by 4.0.
\newblock \urlprefix\url{https://www.iea.org/reports/tracking-clean-energy-progress-2023}.

\bibitem[{Tsujimura and Suzuki(2017)}]{Tsujimura.2017}
Tsujimura, T. and Suzuki, Y. (2017).
\newblock The utilization of hydrogen in hydrogen/diesel dual fuel engine.
\newblock \emph{International Journal of Hydrogen Energy}, 42(19), 14019--14029.
\newblock \doi{10.1016/j.ijhydene.2017.01.152}.

\end{thebibliography}

% \appendix

\end{document}